\documentclass{sigchi}

\toappear{
Permission to make digital or hard copies of all or part of this work
for personal or classroom use is granted without fee provided that
copies are not made or distributed for profit or commercial advantage
and that copies bear this notice and the full citation on the first
page. Copyrights for components of this work owned by others than ACM
must be honored. Abstracting with credit is permitted. To copy
otherwise, or republish, to post on servers or to redistribute to
lists, requires prior specific permission and/or a fee. Request
permissions from \href{mailto:Permissions@acm.org}{Permissions@acm.org}. \\
\emph{CHI '16},  May 07--12, 2016, San Jose, CA, USA \\
ACM xxx-x-xxxx-xxxx-x/xx/xx\ldots \$15.00 \\
DOI: \url{http://dx.doi.org/xx.xxxx/xxxxxxx.xxxxxxx}
}

\usepackage{balance}       
\usepackage{graphics}      
\usepackage[T1]{fontenc}   
\usepackage{txfonts}
\usepackage{mathptmx}
\usepackage[pdflang={en-US},pdftex]{hyperref}
\usepackage{color}
\usepackage{booktabs}
\usepackage{textcomp}

\usepackage{microtype}        
\usepackage{ccicons}          

\usepackage{todonotes}

\def\plaintitle{OralCam: Enabling Self-Examination and Awareness of Oral Health Using a Smartphone Camera}

\def\emptyauthor{}
\def\plainkeywords{Oral health; Mobile health; Artificial intelligence; Deep learning.}

\makeatletter
\def\url@leostyle{%
  \@ifundefined{selectfont}{
    \def\UrlFont{\sf}
  }{
    \def\UrlFont{\small\bf\ttfamily}
  }}
\makeatother
\def\pprw{8.5in}
\def\pprh{11in}

\definecolor{linkColor}{RGB}{6,125,233}
\hypersetup{%
  pdftitle={\plaintitle},
  pdfauthor={\emptyauthor},
  pdfkeywords={\plainkeywords},
  pdfdisplaydoctitle=true, 
  bookmarksnumbered,
  pdfstartview={FitH},
  colorlinks,
  citecolor=black,
  filecolor=black,
  linkcolor=black,
  urlcolor=linkColor,
  breaklinks=true,
  hypertexnames=false
}

\begin{document}
\title{\plaintitle}

\numberofauthors{1}
\author{
A preprint \\ 
  \alignauthor Yuan Liang$~^{1}$, Hsuan-Wei Fan$~^{2}$, Zhujun Fang$~^{3}$,
Leiying Miao$~^{4}$, Wen Li$~^{4}$, \\ Xuan Zhang$~^{4}$, Weibin Sun$~^{4}$,
Kun Wang$~^{1\textsuperscript{*}}$, Lei He$~^{1\textsuperscript{*}}$, Xiang `Anthony' Chen$~^{1\thanks{Corresponding authors.}}$ \\
    \affaddr{$^{1}$~UCLA, $^{2}$~Tsinghua University, $^{3}$~UCD, \\ $^{4}$~Nanjing Stomatological Hospital, Meidcal School, Nanjing University} \\
    \email{liangyuandg@ucla.edu, fanxw16@mails.tsinghua.edu.cn, zhfang@ucdavis.edu, miaoleiying80@163.com, 18844501367@163.com, zhxuan2015@163.com, wbsun@nju.edu.cn, wangk@ucla.edu, lhe@ee.ucla.edu, xac@ucla.edu}
}

\maketitle

\begin{abstract}
Due to a lack of medical resources or oral health awareness, oral diseases are often left unexamined and untreated, affecting a large population worldwide.
With the advent of low-cost, sensor-equipped smartphones, mobile apps offer a promising possibility for promoting oral health. 
However, to the best of our knowledge, no mobile health (mHealth) solutions can directly support a user to self-examine their oral health condition.
This paper presents \textit{OralCam}, the first interactive app that enables end-users' self-examination of five common oral conditions (diseases or early disease signals) by taking smartphone photos of one's oral cavity.
\textit{OralCam} allows a user to annotate additional information (\textit{e.g.} living habits, pain, and bleeding) to augment the input image, and presents the output hierarchically, probabilistically and with visual explanations to help a laymen user understand examination results.
Developed on our in-house dataset that consists of 3,182 oral photos annotated by dental experts, our deep learning based framework achieved an average detection sensitivity of 0.787 over five conditions with high localization accuracy. 
In a week-long in-the-wild user study (N=18), most participants had no trouble using \textit{OralCam} and interpreting the examination results.
Two expert interviews further validate the feasibility of \textit{OralCam} for promoting users' awareness of oral health. 
\end{abstract}




\section{1. Introduction}
Oral disease is a major and growing global public health challenge.
Without being treated properly, it can lead to individual's pain, impairment of function, reduced quality of life, and added economic burdens \cite{petersen2005global}. 
Early detection of potential oral diseases is important, since it enables interventions that alter its natural course and preventing the onset of adverse outcomes with a minimized cost \cite{deep2000screening}. 
Routine dental visit is the most effective way for oral disease detection \cite{cohen2013expanding}. 
However, due to the lack of dental care resources and awareness of oral health, many oral health issues are often left unexamined and untreated, affecting about 3.5 billion people worldwide according to an estimation in \cite{kassebaum2017global}. 

\begin{figure} 
\centering
  \includegraphics[width=0.85\columnwidth]{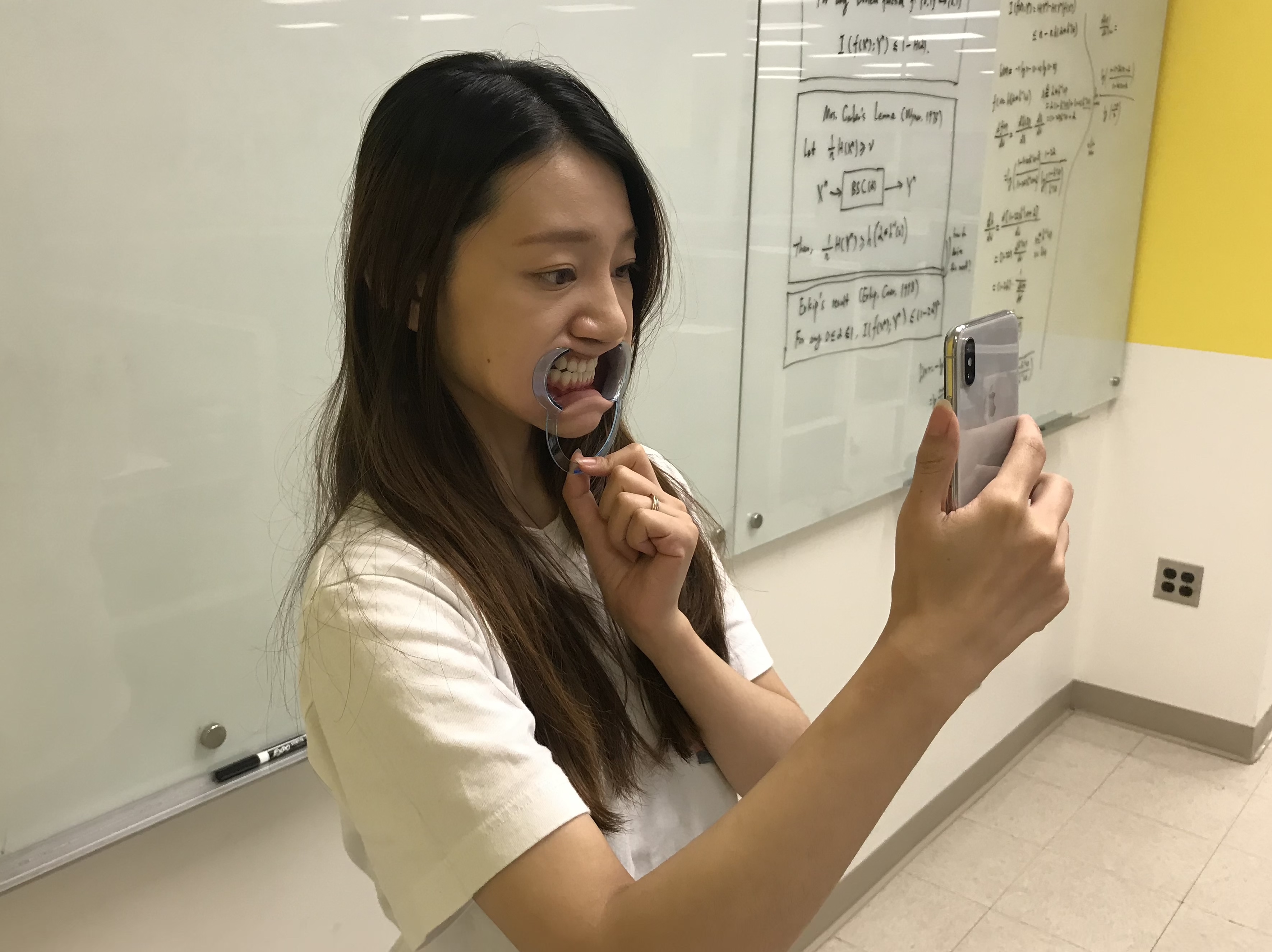}
  \caption{An end-user taking oral cavity photos for self-examination with a smartphone and a dental mouth opener.}
  \label{take_photo}
  \vspace{-7mm}
\end{figure}

As smartphones become increasingly low-cost and ubiquitous, mobile apps provide a promising solution to detect health issues for everyday users as a supplement to clinical visits. 
Indeed, recent development in mobile health (mHealth) has witnessed the use of smartphone for detecting hypertension \cite{wang2018seismo}, liver disorder \cite{mariakakis2017biliscreen}, traumatic brain injury \cite{mariakakis2017pupilscreen}, skin issues \cite{vardell2012visualdx}, and evaluating the overall health condition \cite{ding2019reading}.
However, to the best of our knowledge, the use of smartphone for detecting oral conditions is still an underexplored area. 
As shown in a recent study \cite{tiffany2018mobile}, although more than 1,075 oral health related apps are publicly available, none of them enables directly examining an user's oral health condition.
To bridge this gap, we propose to enable self-examination of oral health by capturing images of oral cavity with a smartphone's camera, and automatically detect potential oral conditions with a data-driven model. 
Moreover, specific preventative actions can be suggested to users to improve their oral health \cite{nasseh2014effect}.  

To inform the design of self-examining oral conditions, we interviewed three dental experts and identified five common yet easily ignored oral conditions our system should support: periodontal disease, caries, soft deposit, dental calculus, and dental discoloration. 
Further, we formulated four categories of system requirements from a clinical point of view: {\em (i)}~ accurate automatic assessment, {\em (ii)}~ leveraging additional modalities besides images, {\em (iii)}~ providing contextual results, and {\em (iv)}~ suggesting targeted actions according to examination results.

We then developed \textit{OralCam}---the first mobile app that supports self-examination and awareness of oral health with automatic detection of oral conditions. 
\textit{OralCam} is embedded with an image-based deep learning (DL) model. To train the model, we  built an in-house dataset that consists of 3,182 oral cavity images with detailed annotations from dental experts.  
In terms of user input, \textit{OralCam} goes beyond existing approaches that purely rely on the image modality \cite{jiang2017artificial,mariakakis2017pupilscreen,mariakakis2017biliscreen}; 
instead, we allow users to provide additional information (\textit{e.g.} living habits, history of pain and bleeding) via a survey and annotating the captured images, which serves as model priors to enhance its performance. 
In terms of system output, \textit{OralCam} goes beyond prior work that conveys DL results to physicians \cite{mariakakis2017biliscreen,ding2019reading,xu2018ecglens}; 
instead, \textit{OralCam} targets at laymen users by showing exam results hierarchically to avoid information overload, and contextualizing each identified condition with textual description of probability. 
Moreover, \textit{OralCam} not only identifies the existence of oral issues, but also informs the user where the model is looking at by highlighting related regions on input photos with a heatmap or bounding boxes. 

We conducted three evaluations to validate \textit{OralCam}:
{{\em (i)}~} A technical evaluation of model accuracy shows our model achieves an average classification sensitivity of 0.787 and accurate condition localization verified with dentists. 
{{\em (ii)}~} A user evaluation with 18 participants over a one-week period indicates that our app enables self-examination, can effectively present exam results, and improves users' understanding of oral health. 
{{\em (iii)}~} An expert evaluation with 2 board-certified dentists suggests that our data collection mechanism is clinically valid, lighting and focus mainly cause model's inaccuracy, and the results can promote users' awareness of their oral health. 

\subsection{Contributions}
Our main research contributions include:
\begin{itemize}
    \item \textit{OralCam}---the first mobile app that enables self-examination and awareness of five common oral conditions;
    \item End-to-end pipeline---processing mixed-modality images with priors as input to classify and localize oral conditions;
    \item Evaluation---a user study and an expert study validating the feasibility of enabling end-users to independently self-examine oral condition via a personal smartphone.
\end{itemize}

\begin{figure*} 
\centering
  \includegraphics[width=0.8\textwidth]{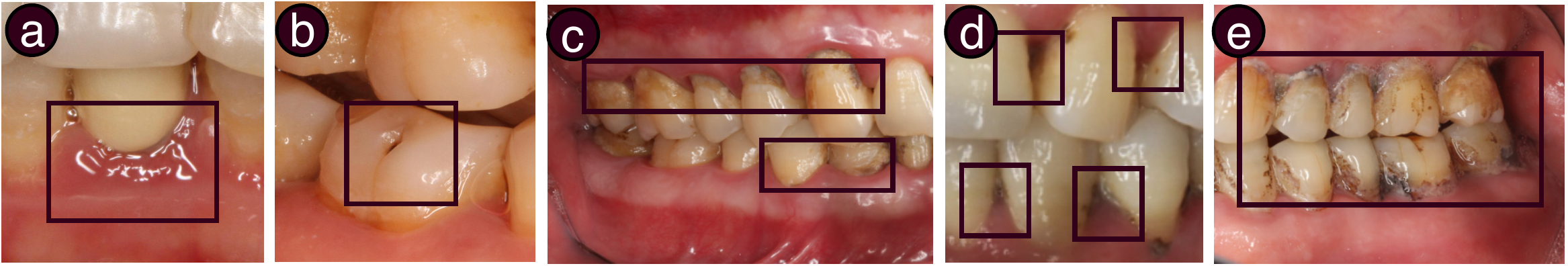}
  \caption{Oral cavity images that contain oral conditions of: (a) periodontal disease, (b) caries, (c) soft deposit, (d) dental calculus, and (e) dental discoloration. Typical areas of the conditions are highlighted with black boxes. 
  }
  \label{typical_path}
  \vspace{-5mm}
\end{figure*}

\section{2. Background of oral conditions}

To set the scenario, we provide a brief overview of the five oral conditions for which \textit{OralCam} supports a user to perform self-examination. 
We have collaborated with three dental experts to identify these five conditions, all of which are {\em (i)}~ common in the general population yet often easily ignored, {\em (ii)}~ vital not just to dental but the overall health of the entire body, and {\em (iii)}~ visually diagnosable using images of one's oral cavity. 
Typical appearance of the five conditions is shown in Figure \ref{typical_path}, highlighted with black boxes. 

\textbf{Periodontal disease} is one of the most prevalent untreated oral diseases \cite{nazir2017prevalence}. 
It is a chronic inflammation of the gum, and can lead to tooth loss at an advanced stage \cite{de2009periodontitis}. 
While clinical operations are needed for treatment, healthy lifestyles and proper hygiene routines help to relieve and prevent \cite{eke2012prevalence}.

\textbf{Caries} is damage of a tooth due to acids made by bacteria, which affecting 43.1\% of the population aged 2-19 in the U.S. \footnote{https://stacks.cdc.gov/view/cdc/53470}.
Untreated caries can cause pain, infection, and affect the overall health. 
Proper intervention in the continuing process of caries can stall or reverse its progress \cite{featherstone2008dental}.

\textbf{Soft deposit} is a visible build-up of bacteria on teeth and a major cause for dental decay and gum inflammation \cite{broadbent2011dental}. 
Untreated soft deposit can lead to dental calculus and periodontal diseases \cite{mariotti1999dental}. 
Soft deposit can be controlled or removed via proper hygiene routines.

\textbf{Dental calculus} is mineralized soft deposit that damages periodontal tissues by adsorbing toxic products \cite{ash1964correlation}. 
It is a major etiological factor of periodontal disease. 
While proper hygiene routines can control its development, dental calculus can only be treated in clinical settings \cite{jepsen2011calculus}.

\textbf{Dental discoloration} can arise from many lifestyle-related factors, \textit{e.g.} tobacco use, diet, and personal hygiene \cite{hattab1999dental}.
As reviewed in \cite{heymann2005tooth}, many prevention and treatments are available for dental discoloration, including proper hygiene routines.

\section{3. Related work} 
In this section, we first show recent mHealth apps using smartphone cameras for health sensing, and then focus on current technologies for oral health. 
We also summarize the DL-based diagnostic algorithms that related to our detection model.  

\subsection{Health Sensing with Smartphone Cameras}
Smartphones have become ideal platforms for always-available health sensing, since nowadays they are usually empowered with various sensors. 
Importantly, smartphone cameras with enhanced resolution have enabled many health sensing applications.
For example, Ding \textit{et al.} \cite{ding2019reading} developed an everyday health monitoring app by using face photos;
BiliScreen \cite{mariakakis2017biliscreen} detected live disorders by capturing Jaundice color changes from smartphone photos;
Vardell \cite{vardell2012visualdx} introduced a skin disease detection system with skin photos as a source of input. 
For \textit{OralCam}, we utilize smartphone cameras to capture high quality photos of users' oral cavity as important inputs for automatic oral condition examination.

\subsection{Oral Health Related Technologies}
HCI community has shown constant interests in exploring technologies for promoting oral health in recent years. 
Most of the existing studies focus on tracking and improving tooth hygiene behaviours. 
For tracking hygiene behaviours, researchers have employed various techniques with optical motion sensors \cite{chang2008playful}, inertial sensors \cite{huang2016toothbrushing,akther2019moral}, smartphone cameras \cite{yoshitani2016lumio}, and more built-in sensors \cite{korpela2015evaluating,ouyang2017asymmetrical}.  
For improving hygiene behaviours, Chang \textit{et al.} \cite{chang2008playful} introduced a play-based virtual education system; Text2Floss \cite{hashemian2015t} and Brush DJ \cite{underwood2015use} studied the effectiveness of mobile apps for motivating hygiene behaviours. 

Numerous oral health related apps have also emerged from mobile app stores. 
Recent work \cite{tiffany2018mobile,parker2019availability} has reported the state of publicly available oral health apps by comprehensively reviewing the content of the most popular ones. 
These apps have a variety of functions, \textit{e.g.} promoting hygiene behaviours, providing oral health education, and motivating healthy lifestyle \cite{tiffany2018mobile}.
However, none of these apps have considered the oral health condition of a user in their functions. 
We argue that the awareness of a user's oral health condition can improve the functionality of an app, since the presented contents can be tailored based on the individual's situation. 
Moreover, users might be more motivated to follow interventions when being informed about their own oral issues \cite{clawson2015no}.

To conclude, there is no existing mHealth solution to maintain an awareness of a user's oral conditions. 
Different from existing work, \textit{OralCam} enables self-examination of oral health for everyday usage, which is complementary to clinical visits. 

\subsection{Diagnostic DL Algorithms}
Recently, Deep Convolutional Neural Network (DCNN) algorithms became the state-of-the-art solutions for many diagnostic tasks with medical images \cite{litjens2017survey,esteva2019guide}. 
Two types of diagnostic tasks are related to our work: {\em (i)}~ image classification, which predicts whether some findings exist in an image, and {\em (ii)}~ object localization, which further localizes the findings with bounding boxes. 
For the former task, a DCNN model first performs feature extraction by iteratively warping the image through convolutional operations, and then transforms the features into a probability distribution over target categories \cite{esteva2017dermatologist,cheng2016computer}; 
for the latter task, a DCNN model follows the same feature extraction step, and then regresses features for coordinates of bounding boxes \cite{de2018clinically,cicero2017training}. 

Moreover, Multi-task Learning (MTL) has recently been developed for cases where multiple types of tasks are simultaneously required of a model \cite{ranjan2017all}. 
Specifically, MTL uses one DCNN model with a shared feature extraction step and separate regression steps to solve all types of tasks in a single shot. 
It has been shown that MTL can improve accuracy for each task compared to using separate models \cite{ranjan2017all,xu2018less}, since the extracted features has better generalization through the joint training process.

For \textit{OralCam}, we formulate the detection of different oral conditions based on their specific clinical nature. 
Since each disease can thought of as a separate task, we follow the MTL framework to solve all the tasks one single DCNN model. 
\section{4. System requirements}
To understand the system requirements of \textit{OralCam} from a clinical point of view, we conducted interviews with three dentists from a hospital of stomatology.
We started by describing the motivation and goal of \textit{OralCam}, emphasizing a focus on using smartphone camera to capture and assess oral health conditions. 
We gathered and built on dentists' initial feedback to flesh out a list of system requirements that ensure clinically-valid, patient-friendly data capture and reasoning, which we present below.

\begin{itemize}
\item \textbf{R1. Accurate detection.}
Automatic assessment should be accurate so that users can be informed of possible oral conditions without being misled by false detection. 
Given that automatic detection algorithms have imperfect accuracy, the uncertainty or the level of confidence needs to be clearly conveyed to the user to ensure a comprehensive understanding of their oral conditions.

\item \textbf{R2. Leveraging additional modalities.}
In clinical diagnosis, dentists often rely on more than the visuals of a patient's oral cavity. 
Other factors, \textit{e.g.} oral cleaning routines, smoking habits, and patients' descriptions of symptoms, also contribute to diagnosis and should be taken into consideration in tandem with smartphone-captured images.

\item \textbf{R3. Providing contextualized results.}
It is insufficient to simply show whether a condition exists or not. 
To help users comprehensively understand the self-examination results, it is useful to show the localization of oral diseases as well as related background knowledge on demand, \textit{e.g.} a layman description with exemplar images of symptoms. 
Dentists point out that such contextual information (location, background, \textit{etc.}) can also help gain trust from users. 

\item \textbf{R4. Suggesting preventive actions.} 
Dentists consider it important for the system to suggest oral health enhancing activities based on the examination results.
For example, on the existence of soft deposit, dentists have suggested preventive actions, \textit{e.g.} eating more food containing fiber, brushing teeth twice a day with the Bass Method \cite{joybell2015comparison}, using floss and rinsing after meals.
\end{itemize}

\begin{figure*}[h]
\centering
  \includegraphics[width=0.75\textwidth]{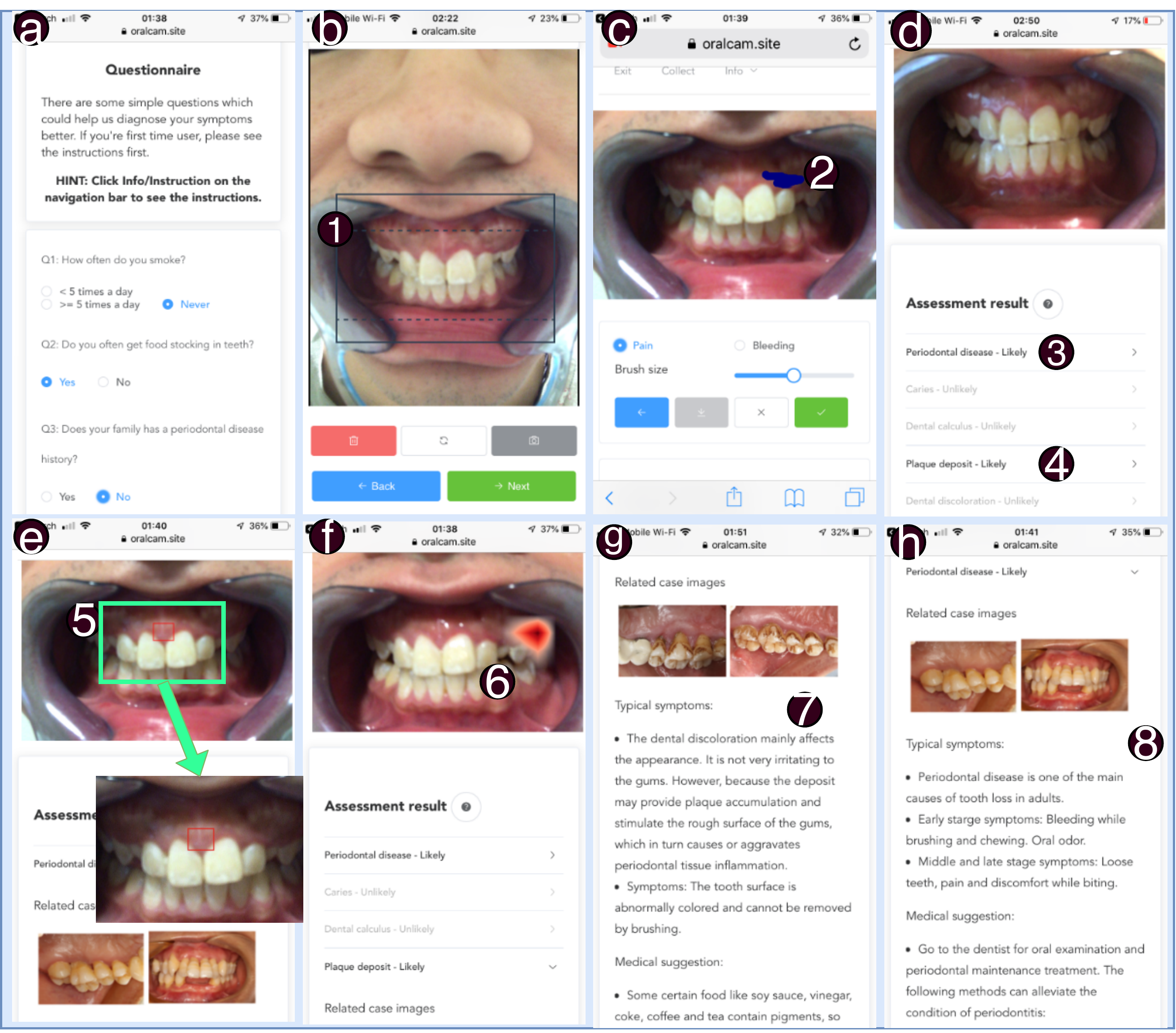}
  \caption{\textit{OralCam} workflow with interfaces.}
  \label{fig_workflow}
  \vspace{-3mm}
\end{figure*}

\section{5. Interaction with OralCam}
Guided by the aforementioned requirements, we designed and implemented \textit{OralCam} for self-examination of oral health. 
The only accessory required for using \textit{OralCam} is a dental mouth opener for opening up one's oral cavity, as shown in Figure \ref{take_photo}, which usually costs less than \$1 \footnote{https://www.amazon.com/s?k=Dental+Mouth+Opener} and can be reusable. 
The flow of interaction using \textit{OralCam} consists of two main steps: 
{\em (i)}~ Input---collecting information that includes oral cavity images and additional symptomatic annotations on the images (\textit{e.g.} regions of pain and bleeding), and
{\em (ii)}~ Output---viewing and exploring examination results. 
Figure \ref{fig_workflow} shows the overall workflow of \textit{OralCam}.

\subsection{Input: Capturing Oral Images with Symptom Annotations}
In terms of collecting information, users start by answering a questionnaire (Figure \ref{fig_workflow}a) of seven multiple-choice questions about hygiene habits and medical history.
Importantly, these questions are commonly used for diagnosis in clinical setting \cite{loesche2001periodontal,hattab1999dental} and users' responses to these questions serve as model priors to improve its performance beyond only relying on images only (\textbf{R2}).  
We describe details of incorporating these questions in the implementation section.
Then, users are prompted to take one or multiple photos of their oral cavity after putting on a mouth opener (Figure \ref{fig_workflow}(b)). 
Taking multiple photos from different angles allows for focusing on various locations in the oral cavity to achieve a comprehensive examination. 
Photo taking is guided with a dashed box (Figure \ref{fig_workflow}(1)),
which helps a user position to align the camera for a clear view of the oral cavity. 
The photo will later be cropped by the solid box (Figure \ref{fig_workflow}(1)) to a aspect ratio and reception region similar with training images, in order to relieve the performance degradation from image appearance shift (\textbf{R1}). 
Lastly, users can provide additional descriptions of other symptoms (\textit{e.g.} pain and bleeding) by directly drawing on the captured images, which are then incorporated into the model as feature maps to enhance its performance (\textbf{R2}), as detailed in the implementation section. 

\subsection{Output: Viewing \& Exploring Self-Examination Results}
\textit{OralCam} presents examination results hierarchically, probabilistically, and with visualized explanation for layman user to understand their oral conditions.
For each image, \textit{OralCam} outputs the likelihood of having each type of diseases (Figure \ref{fig_workflow}(d)), which is grouped into three levels: {\em (i)}~ unlikely, {\em (ii)}~ likely (Figure \ref{fig_workflow}(4)), and {\em (iii)}~ very likely (Figure \ref{fig_workflow}(3)).
The level is classified by applying pre-defined thresholds to the confidence values of DL model for each condition type, which is detailed later in the implementation section. 
Compared to showing simple binary conclusions \cite{esteva2017dermatologist,mariakakis2017pupilscreen}, \textit{OralCam} conveys richer information from examination: users are prompted to take a close look at confident detections, and are also notified of less confident ones to reduce finding misses (\textbf{R1}). 
Once clicking on a detected condition, \textit{OralCam} expands the disease label to reveal the next hierarchy information, \textit{e.g.} typical appearances of such disease (Figure \ref{fig_workflow}(7, 8)), common symptoms, and backgrounds of the disease. 
All this information serves to contextualize the user's understanding of an oral condition beyond a simple textual label. 
Further, \textit{OralCam} highlights related regions of each detected oral condition on the input image, as shown in Figure \ref{fig_workflow}(e, f). 
For examples, the red box in Figure \ref{fig_workflow}(5) gives exact localization of periodontal diseases; 
and a heatmap in Figure \ref{fig_workflow}(6) give hints on regions related to the dental discoloration, with a higher temperature indicates the stronger spatial relevance. 
By visualizing where the model is `looking at', \textit{OralCam} assists layman users to understand examination results. 
Moreover, such an effort allows users to glance into the process of result generating, which can possibly increase a user's trust of the underlying model (\textbf{R3}). 
Finally, \textit{OralCam} also provides condition-specific suggestions of actions, under each condition that detected. 
We came up with these suggestions by discussing with dental experts (\textbf{R4}). 
Figure \ref{fig_workflow}(g and h) shows the examples of suggestions for dental discoloration and periodontal disease.

\section{6. Implementation} 
In this section, we describe the implementation details of \textit{OralCam}.
We first carefully formulate the oral condition detection as a mix of object localization and image classification, by considering the clinical nature of the five conditions. 
Then we utilize an existing image-based DL framework as baseline model, and further improve it from three aspects:
{\em (i)}~ to improve detection accuracy, we enhance the baseline model with non-image information as priors, 
{\em (ii)}~ to localize regions of conditions, we enable activation map based reasoning for model attention, and 
{\em (iii)}~ to convey confidence information, we carefully design two model operating points for each condition. 
We build an in-house dataset of 3,182 oral images with detailed annotations for model development.
Moreover, we make our detection models publicly available at \url{https://github.com/liangyuandg/MobileDentist} to facilitate further researches. 

\subsection{Problem Formulation}

Our detection model should output: {\em (i)}~ image-wise confidence values for oral conditions' existence, and {\em (ii)}~ locations of the existing conditions.
As such, we formulate the object localization task \cite{de2018clinically,cicero2017training} for periodontal disease, caries, and dental calculus, which outputs object bounding boxes that can be directly used for locations. 
We then determine the image-wise confidence of a condition existing with the highest confidence value of all boxes. 
However, for soft deposit and dental discoloration, the findings usually spread over the whole oral cavity \cite{broadbent2011dental,heymann2005tooth}, which can be clearly seen on Figure \ref{typical_path}(c, e).
Thus, formulating these two conditions as localization can be problematic, 
and leads to heavy workload of data annotating for supervised training. 
As a result, we formulate image classification tasks \cite{esteva2017dermatologist,cheng2016computer} for soft deposit and dental discoloration, while their regions of interest are reasoned as heatmaps based on model attention, which will be described in a following subsection.

\subsection{Baseline Model: Input Images Only}
\begin{figure}
\centering
  \includegraphics[width=\columnwidth]{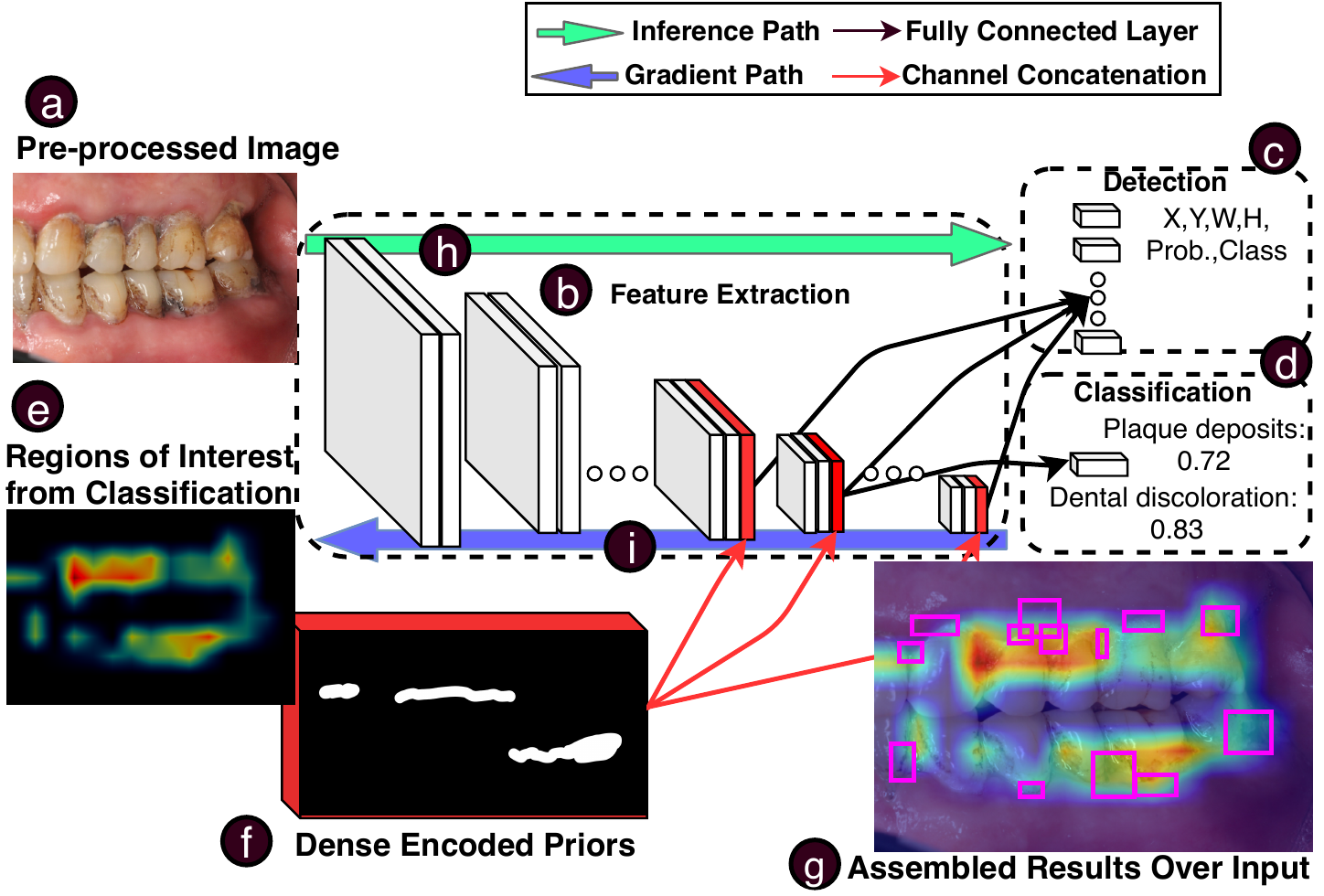}
  \caption{Overview of the DCNN model for oral condition detection.
  }\label{figure_model}
  \vspace{-5mm}
\end{figure}

Leveraging the recent development in DL, we employ a DCNN-based model. 
Moreover, since there are types of tasks (\textit{i.e.} image classification and object localization), we utilize the MTL \cite{xu2018less,ranjan2017all} framework to solve all tasks as a unified model to reduce model redundancy. 
Figure \ref{figure_model} shows the overview of this model.

Our baseline model only takes oral cavity images (Figure \ref{figure_model}(\textbf{a})) as input, with pre-processing steps including spatial normalization and color normalization \cite{redmon2017yolo9000}. 
The deep features of the input image are extracted through a stack of convolutional and non-linear operations (Figure \ref{figure_model}(\textbf{b})), which are trained to be discriminative for detection with back-propagation. 

For image classification tasks, we define the output as a vector $y$ of length 2, whose values represent the confidence scores of containing soft deposit and dental discoloration.
The vector $y$ is derived from regression of the feature map as fully connected layers, as shown in Figure \ref{figure_model}(\textbf{d}). 
For localization task output, the output is defined as a set of vectors of length 5.
Each vector encodes one detected bounding by its coordinates, height, width, and confidence score. 
Similarly, the vector set is derived from the regression of feature maps as fully connected layers, as shown in Figure \ref{figure_model}(\textbf{c}).

While the baseline model only utilizes an image for detection, experienced dentists often inquire for additional information, \textit{e.g.} hygiene habits, medical history, and other signs and symptoms. 
Motivated by this, we propose to leverage additional user-reported information to augment the input images for improved accuracy. 

\subsection{Enhanced Model: Input Images + Priors}
Computationally, fusing additional information such as priors for image-based DCNN is promising \cite{esteva2019guide}.
Previous work has mainly focused on utilizing Electronic Health Record (EHR) as a source of priors \cite{zhang2017mdnet,wang2018tienet}.
However, since EHR is usually written in natural languages without standard structures \cite{reisman2017ehrs}, it is challenging for extracting information, and further, training the model.
To circumvent these limitations, \textit{OralCam} administers multiple-choice questionnaire to collects a user's oral health history, smoking habits and oral cleaning routines (Figure \ref{fig_workflow}(a)), which provides structured data. 
Together with user-drawings of pain and bleeding regions (Figure \ref{fig_workflow}(b)), which are also structured, the information can be incorporated into the model as priors. 

Specifically, we one-hot encode answers to the questionnaire as a feature vector and drawings as a feature map. 
We further replicate the feature vector spatially to make it the same size as the input image, and stacked it with the feature map as in \ref{figure_model}(\textbf{f}).
Then we incorporate the encoded information to the regression layers of the baseline model by feature-wise concatenating with the deep feature maps from DCNN, as shown with the red arrows in Figure \ref{figure_model}.
Thus, the model derives the outputs of classification and localization from regression over both DCNN features and priors. 
Moreover, added complexity of the enhanced model is minimal, since the model architecture remains the same, and the only parameters affected are the kernels at the last layers before model output. 
As thus, we derive the enhanced model from the baseline model by fine-tuning only the regression parameters to leverage the learnt features. 
Experiments (detailed later) show that the enhanced model with priors achieves improved accuracy than the baseline model that relies on input images alone. 

\subsection{Presentation of Examination Results}
For localization tasks, examination results are shown as bounding boxes that encompass regions where specific oral conditions occur. 
For classification tasks, we utilize the Classification Activation Mapping (CAM) technique \cite{zhou2016learning,selvaraju2017grad} to visualize the spatial attention of the model when classifying the input image. 
Specifically, we implemented Gradient-weighted CAM \cite{selvaraju2017grad}, which achieves accurate localization for \textit{OralCam} as reported later in our experiments.
As shown with the blue arrows (Figure \ref{figure_model}(\textbf{i})), the gradient of a classification output $y$ is propagated backwards, producing a heatmap (Figure \ref{figure_model}(\textbf{e})) of the same resolution as the input image.
Following \cite{selvaraju2017grad}, a higher temperature on the heatmap means the region can be interpreted as more related to the detected condition. 
By assembling bounding boxes from localization tasks and the heatmap from classification tasks, Figure \ref{figure_model}(\textbf{g}) shows a schematic drawing with regions of all detected conditions over the input image. 

Moreover, to clearly convey confidence of model on a finding as shown in Figure \ref{fig_workflow} (7 and 8), we set two operating points, named the $1st$ and the $2nd$ operating point, for each condition with different confidence thresholds. 
The $1st$ operating point has a high threshold: any finding (\textit{i.e.} classification or bounding boxes) with a confidence value higher than that indicates the model is confident about it, and we put it as "\textit{very likely}" to users; 
As such, the $2nd$ one has a lower confidence threshold: we put findings with confidence values in between as "\textit{likely}", and those below this threshold as negative, or "\textit{unlikely}".    
This design enables trade-offs between miss rate and false positive \cite{de2009accuracy} for an imperfect model: the $1st$ operating point only highlights confident findings with a lower sensitivity, while the $2nd$ one keeps the miss rate of model low via including findings that are not so confident. 
We report our operating point selection with corresponding model sensitivities and false positive rates in the evaluation part.

\subsection{Dataset and Model Training}
For modeling, we have built an in-house dataset of oral cavity images with detailed annotations from dentists. 
Our dataset consists of 3,182 images collected from around 500 volunteers at a stomatology hospital by 6 dentists, with a wide age coverage from 14 to 60.
We control a balanced number of healthy volunteers and volunteers with conditions, by collecting images from departments of orthodontics, endodontics, and periodontics.
Among the images, 1,744 show periodontal diseases, 441 show discolorations, 1017 show caries, 712 show soft deposits, and show 739 calculus.
Note that each image can show none, one or more types of conditions.

All the images have been annotated for the five oral conditions by three board-certified dentists. 
First, three dentists divided up the 3,182 images, with no overlap, for separate labeling. 
Then, the dentists went through all image-label pairs to reach consensus by discussion. 
The majority voting was used for very few cases (boxes) that could not reach consensus. 
We used different annotation methods for two types of tasks (\textit{i.e.} image classification and object localization): 
periodontal disease, caries, and dental calculus have been annotated with bounding boxes, 
while soft deposit and dental discoloration are annotated with image-level labels.
Moreover, since there are no well-defined boundaries for bounding boxes in our case (and many other medical-related cases), we followed a common approach and instructed the dentists to focus on the correctness of box centers. 
The dataset was randomly divided into a training set of 2,182 images (200 of them for stop criteria), and a testing set of 1,000 images, where we ensured that images of the same person with the same cavity view did not occur in both sets. 
Moreover, there are 293 images augmented with additional information collected from the aforementioned questionnaire and drawings, while 160 of them are within the training set and 133 in the testing set. 

For model training, we follow \cite{xu2018less} by optimizing a loss formulated as the equally weighted sum of localization loss \cite{liu2016ssd} and classification loss \cite{esteva2017dermatologist}. 
We first train our baseline model 
and afterwards, the enhanced model by initializing it from the trained baseline model, and fine-tuning the regression operation parts for incorporating priors.  
Moreover, we employ intensive augmentation to images \cite{redmon2017yolo9000} including spatial shifts (random crops, rotations, and scaling) and color channel shifts (random hue, saturation, and exposure shifts), in order to increase the robustness of our model for in-the-wild application with uncertain factors, \textit{e.g.} camera angles, distances, and lighting conditions. 

\section{7. Evaluation}
We perform three evaluations of \textit{OralCam}: {\em (i)}~ a technical evaluation of model performance on detecting the five oral conditions, {\em (ii)}~ a week-long deployment study from 18 end users, and {\em (iii)}~ an expert interview with two board-certified dentists.

\begin{figure*}
\centering
  \includegraphics[width=\textwidth]{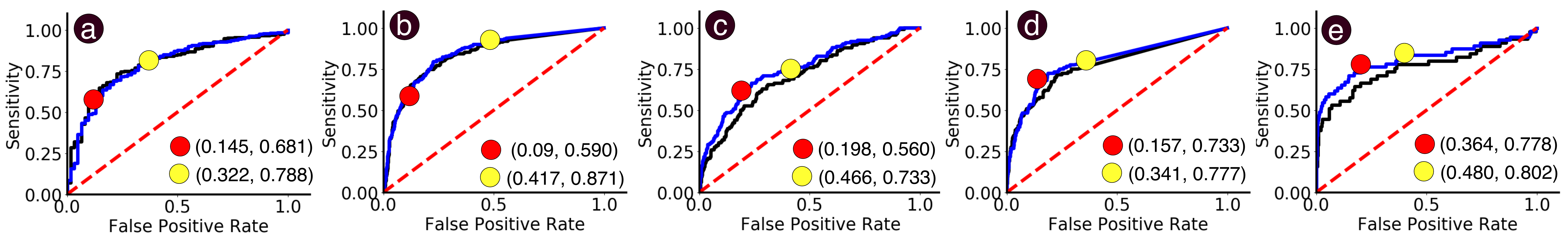}
  \caption{Classification performance for: (a) periodontal disease, (b) caries, (c) soft deposit, (d) dental calculus, and (e) dental discoloration. The black solid lines show the ROC curves of our baseline model, while the blue solid lines for the enhanced model with priors. The red dashed lines show the performance of random chance. The red and yellow dots show our designed $1st$ and $2nd$ operating points, and their performance is shown in the format of (false positives per image, sensitivity).} \label{rocs}
  \vspace{-5mm}
\end{figure*}

\subsection{Technical Evaluation}
We evaluate our detected model from two aspects: {\em (i)}~ classification performance for recognizing the existence of a condition, and {\em (ii)}~ localization performance for pinpointing conditions on images. 
Note that all of our evaluations are based on the data from testing set, which was set apart unseen during the model training. 

In terms of classification performance, we follow \cite{cheng2016computer,esteva2017dermatologist,litjens2017survey} and evaluate our model using the Receiver Operating Characteristic (ROC) curve for all five oral condition, which shows the model sensitivity against its false positive rate with different classifying thresholds.
Figure \ref{rocs} illustrates that both our baseline model (black lines) and enhanced model (blue lines) are capable for accurately telling the existence of five oral conditions, by marginally outperforming the random chance model (red lines). 
Table \ref{tab1} details the area under the curve (AUC) values, which are commonly-used metric for measuring overall classification performance, for the five conditions. 
We can see that our enhanced model achieves an average classification AUC of 0.787 for the five conditions, which boosts the baseline model by 1.55\% with priors incorporated. 
Moreover, we argue that more training data with collected priors can help further improve the performance.

Importantly, we show our selected $1st$ and $2nd$ operating points as red and yellow dots with their performance on the ROC curves in Figure \ref{rocs}. 
In specific, our model achieves an average sensitivity of 0.668, and a false positive rate as low as 0.191 at the $1st$ operating point;
and the average sensitivity can reach 0.794 at the $2nd$ operating point, while having a correspondingly higher false positive rate of 0.405. 
As such, by setting a higher confidence threshold for the $1st$ operating point, we reduce the false positive prediction, and show the detected conditions under this point as "\textit{very likely}";  
and by setting a lower threshold for the $2nd$ operating point, we aim at a low miss rate, and flag the conditions under this point as "\textit{likely}".  

\begin{figure}[h!]
\centering
  \includegraphics[width=0.85\columnwidth]{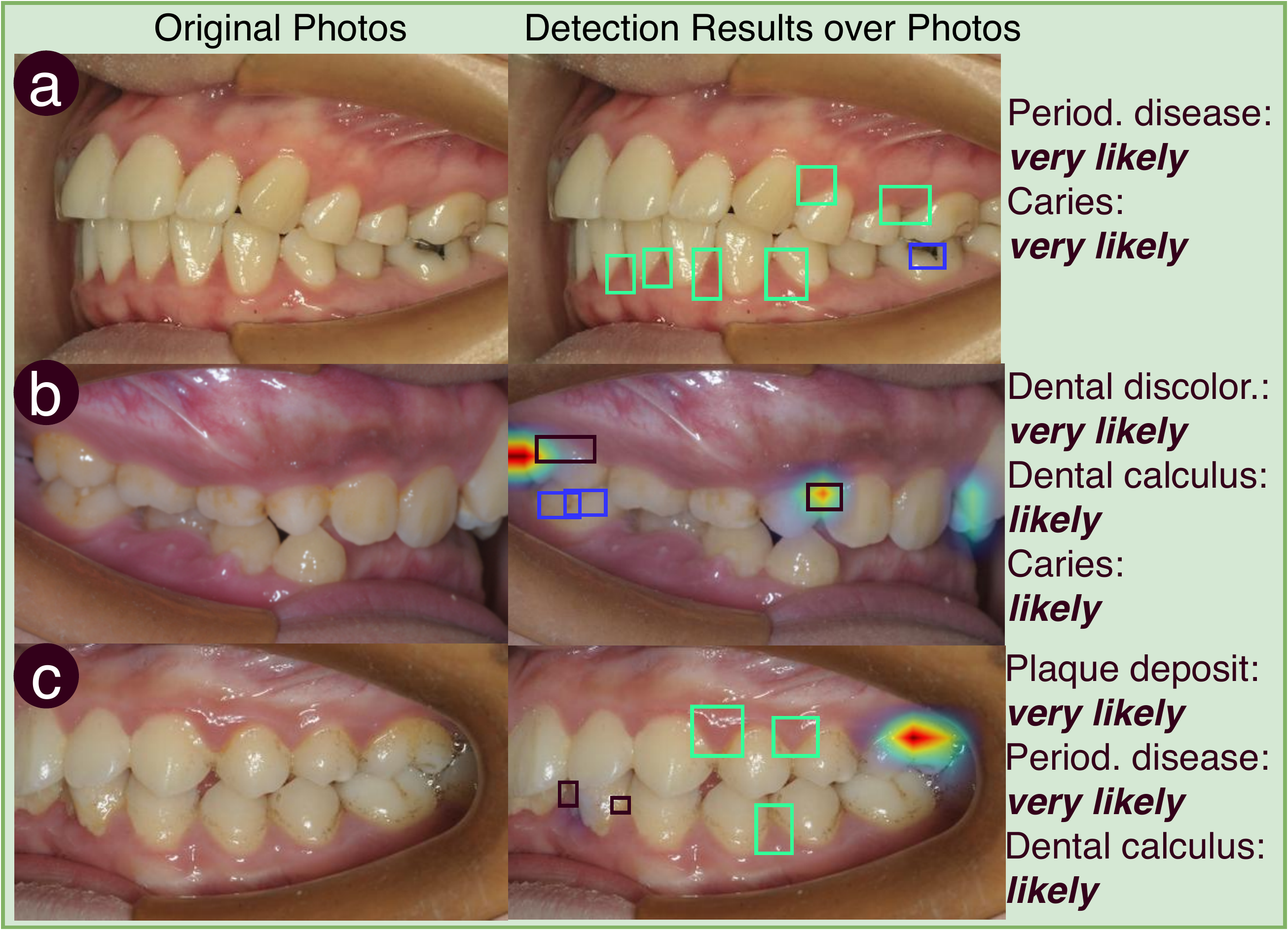}
  \caption{Visualization of detection results from testing data set. The original photos are shown in the left column, while detection results are visualized on the right column. Rows (a), (b), and (c) are three typical oral photos from persons with oral conditions. The rightmost texts describe detection results with confidence levels. Detected periodontal disease, caries, and dental calculus are marked with green, blue and black boxes. Detected dental discoloration (on row (b)) and soft deposit (on row (c)) are marked with heatmap. }
  \label{example_draw}
  \vspace{-4mm}
\end{figure}

In terms of localization performance, we follow \cite{litjens2017survey,dou2016automatic}, and evaluate our model using the Free Response Operating Characteristic (FROC) curve for periodontal disease, caries, dental calculus, since their ground-truth bounding boxes are annotated in the dataset.
In specific, FROC curve shows the box-wise sensitivity against number of false positive boxes per image under different confidence thresholds. 
For soft deposit and dental discoloration where no pixel-level annotations are available, we follow \cite{selvaraju2017grad} by asking two board-certified dentists to give an agreement rating for each localization result. 
Specifically, we show dentists images from testing set that having at least one of the two conditions detected, together with our localization results, which have been visualized in a similar way as in Figure \ref{example_draw}. 
Then, a rating is given to each result on a scale from 1 (strongly disagree) to 5 (strongly agree) by evaluating if the heatmap demonstrates the regions with conditions comparing to dentists' opinion. 

\begin{figure} [h]
\centering
  \includegraphics[width=\columnwidth]{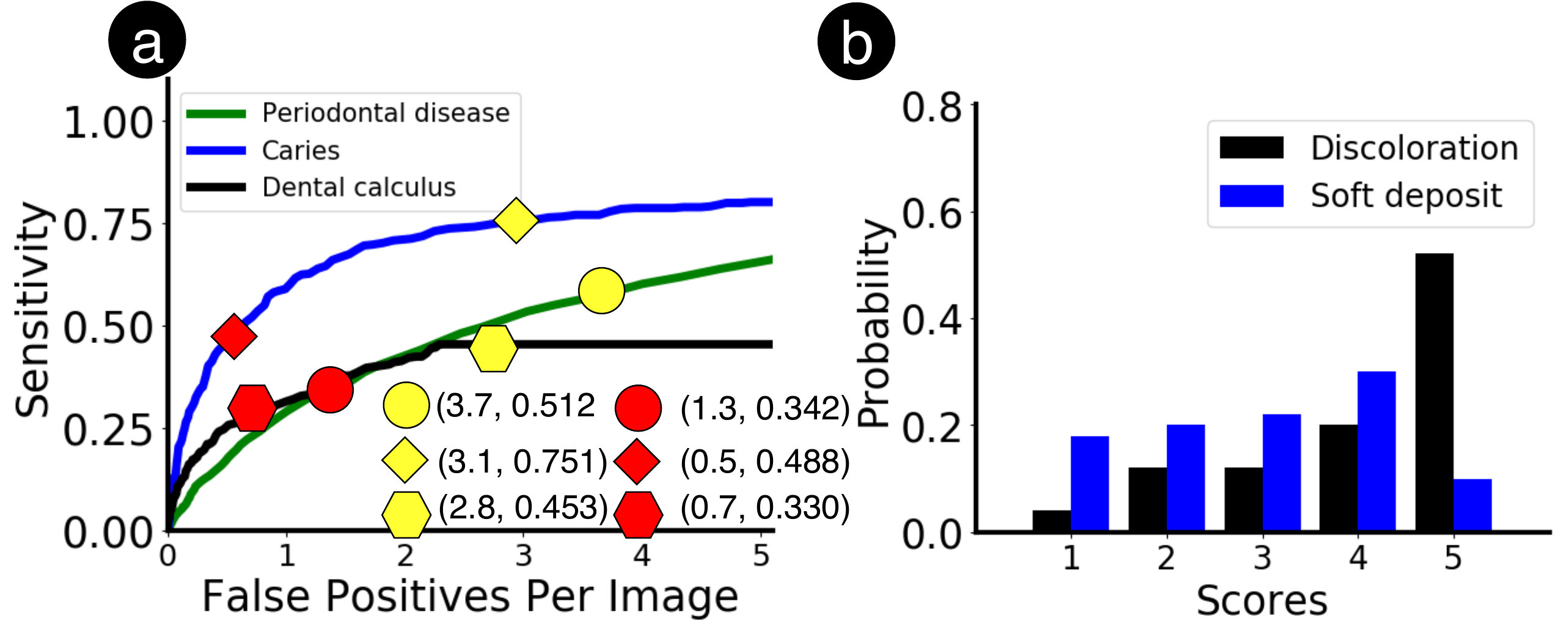}
  \caption{Localization performance of our enhanced model. (a) Green, blue and black lines show the FROC curves for periodontal disease, caries, and dental calculus, respectively. Red and yellow dots on curves show our selected $1st$ and $2nd$ operating points. The performance of operating points is shown in the format of (false positives per image, sensitivity). (b) Distribution of dentists' ratings on the localization for soft deposit and dental discoloration. }
  \label{frocs}
\end{figure}

Figure \ref{frocs} illustrates the capability of our enhanced model for localizing the detected conditions. 
For localization of periodontal disease, caries, and dental calculus, Figure \ref{frocs}(a) shows their FROC curves with our aforementioned operating points. 
In specific, our model achieves an average box-wise sensitivities of 0.387 at the $1st$ operating point, and 0.572 at the $2nd$ operating point. 
For localization of discoloration and soft deposit, Figure \ref{frocs}(b) shows the distribution of dentists' ratings.
In specific, for the two conditions, our model achieves an average rating with medians of 5 and 3, and means of 4.04 and 2.94, all out of 5.  
Overall, the results validate the capability of our model for indicating regions related to all five types of detected conditions.

\begin{table}[h]
\caption{Classification performance of our baseline and enhanced model for periodontal disease (PD), caries (CA), soft deposit (SD), dental calculus (DC), dental discoloration (DD), and their categorized average (Avg.). } \label{tab1}
\centering
\begin{tabular}{l c c c c c c}
\hline
~ & PD & CA & SD & DC & DD & Avg.\\
\hline
Baseline & 0.799 & 0.832 & 0.708 & 0.778 & 0.759 & 0.775 \\
Enhanced & 0.791 & 0.840 & 0.742 & 0.791 & 0.771 & \textbf{0.787} \\
Boost (\%) & -1.00 & +0.96 & +4.80 & +1.67 & +1.58 & \textbf{+1.55} \\
\hline
\end{tabular}
\end{table}

Figure \ref{example_draw} visualizes the output of our model, including both classification and localization, based on three oral photos from the testing set. 
We showcase that our model can effectively detect oral conditions, and indicate regions related to the detected conditions on photos for helping users to understand the results. 

\subsection{End-User Evaluation}
\textit{OralCam} allows a user to self-inspect oral health using a smartphone camera, which we expect to be a nascent area unfamiliar to most mHealth users. 
To validate the feasibility of our approach, we investigate three research questions concerning the process, results and influence of \textit{OralCam}:

\begin{itemize}
    \item {\fontfamily{cmss}\selectfont \textit{RQ1: Process---can users follow instructions provided by \textit{OralCam} to take computationally usable photos of their oral cavity?} } Will users have difficulty aiming the front camera? Can they use the self-reporting tool to describe auxiliary information, \textit{e.g.} pain and bleeding?
    \item {\fontfamily{cmss}\selectfont \textit{RQ2: Results---how do users interpret OralCam's results generated from an image-based oral condition detection model?} } What will users learn from different presentations of \textit{OralCam}'s results---textual summary, annotations, heatmap and comparative examples? Will they trust the results, why and why not?
    \item {\fontfamily{cmss}\selectfont \textit{RQ3: Influence---will users learn about common oral conditions and initiate behavioral changes after using OralCam?} }
\end{itemize}

\textbf{Participants.} To answer these research questions, we conducted a one-week deployment study with 18 participants (aged 21 to 54; 11 male and 7 female). 
Among 18 users, there are 9 engineering students, 4 science students, 3 administration staff, and 2 marketing staff. 
All participants use their own smartphones.

\textbf{Procedure.} The end-user evaluation task consists of the following key activities:

\underline{{\fontfamily{cmss}\selectfont \textit{Introduction.} }}
We kicked off the study by introducing the background and motivation of \textit{OralCam} to each participant. 
Participants then filled out a survey for collecting their demographic and general oral health related information. 
We then demonstrated the smartphone app, let each participant follow a step-by-step onboarding tutorial, and answered their questions in the process. 
After the tutorial, participants were free to continue trying out the \textit{OralCam} until they felt they were able to use it independently.

\underline{{\fontfamily{cmss}\selectfont \textit{`Take-home' study.} }}
Following the introduction, we asked each participant to use \textit{OralCam} as frequently as they would like in the next weeks.
We designed the duration to be one week long since we hypothesized that participants would use \textit{OralCam} at least once a week. 
Participants could access \textit{OralCam} as a web app\footnote{\url{oralcam.site}} via an Internet connection using their own smartphone.

\underline{{\fontfamily{cmss}\selectfont \textit{Exit interview.} }}
At the end of the one-week period, we met with each participant again, surveyed them using the NASA Task Load Index questionnaire, and interviewed them about their experience and reaction of \textit{OralCam} at both an overall level and by a detailed walkthrough of each interactive step.

\begin{figure}
\centering
  \includegraphics[width=0.75\columnwidth]{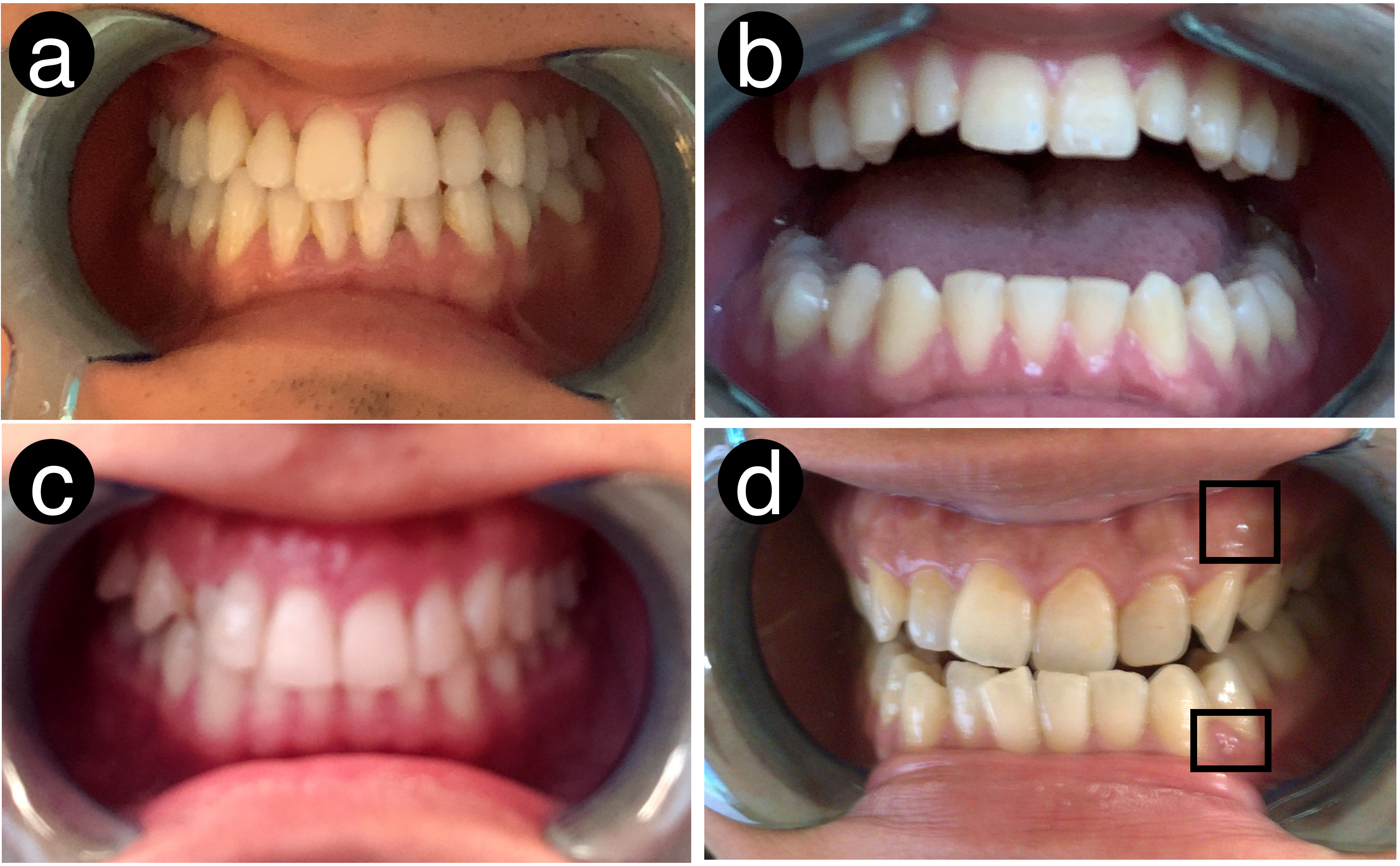}
  \caption{Selected photos taken by users during the in-the-wild study. (a) and (b) are well-captured. (c) has a low resolution since it is wrongly focused. (d) has glare spots because of improper lighting condition, which are highlighted with black boxes.}
  \label{example_cases}
  \vspace{-5mm}
\end{figure}

\textbf{Analysis \& results.}
\underline{{\fontfamily{cmss}\selectfont \textit{Usability.} }}
There were four steps involved when a user provided input to \textit{OralCam}:
(S1) answering the questionnaire, (S2) putting on mouth opener, (S3) taking photos with oral cavity aligning to marked area on screen, and (S4) drawing additional symptoms on images.
We asked users to rate the usability about each of the four steps (from 1--strongly low to 7--strongly high). 
We report the distribution of scores as following: S1 (mean 5.33, median 6, std 1.76); S2(mean 5.94, median 7, std 1.72), S3(mean 4.78, median 5, std 1.62), and S4(mean 5.00, median 6, std 1.60). 
Moreover, figure \ref{example_cases} visualize some selected photos captured by users during the in-the-wild study.  
While the majority of the users reported no problem using \textit{OralCam}, we did notice a few participants who raised issues and concerns.
The survey in S1 was necessary to keep updated information about the user's oral health related information. 
Some participants understood this need and were motivated to answer the survey "\textit{I feel the questions are well designed and can help diagnosis}"(P7). 
At the same time, P13 found the survey is too detailed and assumed some questions, \textit{e.g.} \textit{how many times do you brush teeth a day}, were not related to self-examination.
Regarding S2 and S3, P2 had the most complaint about having to wear a mouth opener; 
P10 found it difficult to target the front camera to oral cavity: "\textit{Personally, taking pictures with the opener is not easy. 
It takes me some time to match the teeth to the specific areas.}"
Similarly, P13 also suggested that removing the requirement to align the oral cavity with the camera would make the process much easier. 
Overall, our information collection design was found usable by the majority of the participants, although improvements can be made to further address the aforementioned complaints. 

\underline{{\fontfamily{cmss}\selectfont \textit{Results Interpretation.} }}
We interviewed our participants about their interpretation of examination results from two aspects: {\em (i)}~ their understanding of the results, and 
{\em (ii)}~ their trust of the results. 
Regarding understanding, every user was asked to describe their conditions if anything was detected, and we checked if the descriptions were consistent by referring to the results.  
We found that the users had no trouble understanding the results, which include the confidence level, condition visualization, and suggestions. 
For example, P15 mentioned the detected soft deposit as: "\textit{it (\textit{OralCam}) says I likely to have some deposit at this left upper teeth. I know I possibly have soft deposit issue, since my dentist has told me before.}"

Regarding the trust, we find that most of participants believe the result: mean and median of scores are 4.94 and 5 out of 7, while 17 participants gave a score higher than 4. 
Noticeably, 12 participants believed they were having some oral conditions that they were not aware of or could not confirm before using the \textit{OralCam}. 
For example, P15 mentioned that he noticed some signs of calculus, and stated that "\textit{I guess I have that (dental calculus). The results confirm my suspicion, so I trust it.}"
This indicates that OralCam is able to communicate oral health related results to laymen users and enable their awareness of their own oral health.
For participants giving a low trust score, some pointed out that without the confirmation of dentist or other kinds of validation, they still remain incredulous about the results.
As an example, P15 mentioned: "\textit{It is better to get support from some authentic institution, like FDA (U.S. Food and Drug Administration) or a certified dentist.}"
We also find out that users can have marginally higher or lower trust when the detection results confirms or conflicts their beliefs. 
For example, P3, detected to have periodontal disease, did not trust the result at all: "\textit{I do not really trust the result. I visit the dentist regularly and I have a pretty good oral hygiene}." 
Others, P17 as an instance, highly trusted \textit{OralCam} because the results validate their previous diagnoses given by their dentists.

\underline{{\fontfamily{cmss}\selectfont \textit{Influence on Awareness and Behaviours.} }}
We first asked participants to evaluate their understanding about the five oral conditions by giving a rating (from 1--strongly low to 7--strongly high) both before and after using \textit{OralCam}. 
We found that the knowledge improvement is significant among participants (p<0.001), although the extent of increment is limited (mean: 1.72, median:1.5). 
Moreover, we were informed by the participants that the major reason for limited improvement is that participants only paid attention to the information of one condition when it was detected by \textit{OralCam}, although such information is available in the app for all five oral conditions. 

Furthermore, 13 participants showed the tendency of using \textit{OralCam} regularly by scoring the willingness as 5 or above on a scale of 7. 
They agreed that using \textit{OralCam} is more convenient than visiting a dentist: taking photos at home and get the results immediately are much easier than attending an dental appointment. 
Regarding using habits, the actual usage frequency during the week among users was: mean 2.1, median 2.0, min 1, max 4, and std 1.1. 
We also inquired for their expected frequency of using.
The answers ranged among once a week (3P), once per two weeks (2P), once per month (2P), and twice a year (1P), and others claimed that it would depend on moods or feelings of oral condition.  
We suggest that further designs of such oral health apps should include suggested usage frequency by working with experts.  

To understand \textit{OralCam}'s influence on users' behaviours, we also asked them if they had changed their cleaning habit (\textit{i.e.} brushing method, frequency, and flossing or not), and if they were likely to change it.
It comes out that only 1 participant (P6) changed the brushing behaviours, by lengthening the time as suggested by the app.
At the same time, however, 10 participants mentioned they were likely to change the behaviours, by giving score of willingness as 5 or above out of 7, which shows the gap between changing awareness and behaviours.  
Moreover, P1 suggested that: "\textit{it helps if it (\textit{OralCam}) is an (offline-based) app, so that it can remind me about brushing method. Then I probably will put it into action.}."

\subsection{Expert Evaluation} 
To clinically validate \textit{OralCam}, we interviewed 2 board-certified dentists (E1: 5 years of practice; E2: 4 years of practice. Both have expertise in periodontal disease treatment). We investigate the following questions:

\begin{itemize}
    \item {\fontfamily{cmss}\selectfont \textit{RQ4: Is our data collection mechanism clinically valid?} }---including patient information, oral images, and pain/bleeding labels.
    \item {\fontfamily{cmss}\selectfont \textit{RQ5: Are the model's results accurate?} }---based on experts' clinical opinion and givn that photos were captured in-the-wild.
\end{itemize}

\textbf{Procedure.} We asked each dentist to review one trial of data (2 images per trial on average) randomly selected from each of the 18 participants in the previous study. 

To answer RQ4, we demonstrated and provided a tutorial of \textit{OralCam} to each expert---similar to the introduction session in the previous user study---followed by an short interview asking experts to comment on the clinical validity of our data collection mechanism.

To answer RQ5, we asked each expert to rate their agreement (from 1--strongly disagree to 7--strongly agree) with the model results for each type of conditions. When leaning towards disagreement, we also asked experts to write a short memo of their own diagnosis. 

\begin{figure}[h]
\centering
  \includegraphics[width=0.75\columnwidth]{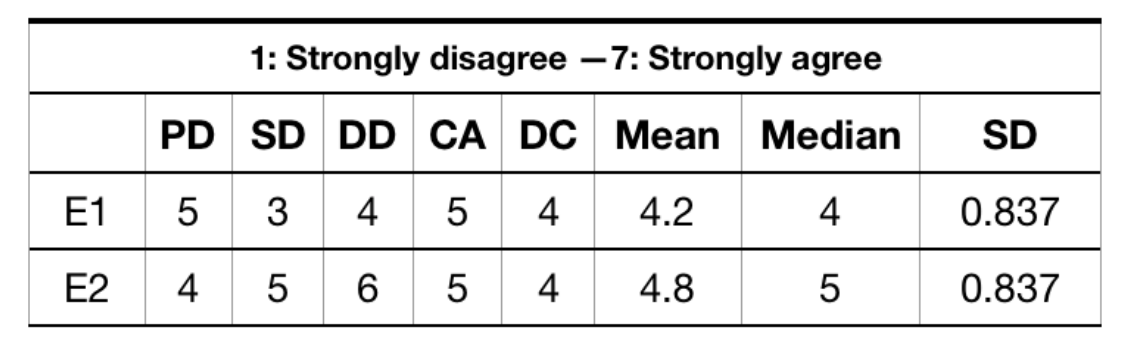}
  \caption{Experts' degrees of agreement with model results for periodontal disease (PD), soft deposit (SD), dental discoloration (DD), caries (CA), and dental calculus (DC). E1 and E2 represent for expert 1 and 2, respectively. }
  \label{user_study_2}
\end{figure}

\textbf{Analysis \& results.} 

\underline{{\fontfamily{cmss}\selectfont \textit{Information Collection Mechanism.} }}
Both experts agreed that our data collection methods are clinically valid.
Moreover, E1 considered the questionnaire very helpful for oral condition detection, and suggested adding more symptom-related questions
\textit{i.e.} feeling teeth shaking.
E2 also agreed on the importance of questionnaire and further stated: "\textit{Questions [in the questionnaire] themselves can be educationally meaningful... If you ask users whether you bleed when eating an apple, they may probably realize this is an abnormal issue}."

Experts observed that two problems of photos lead to detection errors: {\em (i)}~ existence of strong light spots, and {\em (ii)}~ incorrect focus. 
E2 pointed out a detected bounding box was mistakenly placed as "periodontal disease", possibly because there is a light reflection at the location. 
Regarding the focus, E2 noted that a loss of focus could miss certain regions, \textit{e.g.} those of soft deposits
E2 also mentioned that photos with light spots and focus issues might have been taken with artificial lights, where as natural light will provide much better image quality. 
Moreover, both experts mentioned the limitation that the inner cavity cannot be captured due to the lighting condition in the house. 
As E2 mentioned, "\textit{many caries happen on the occlusal surfaces of posterior teeth, and cannot get detected in camera images}."

\underline{{\fontfamily{cmss}\selectfont \textit{Model Accuracy.} }}
The experts rated our detection results of users' photos based on the overall performance of classification and condition localization. 
Figure \ref{user_study_2} shows the experts ratings for the five types of conditions. 
Both experts agreed that \textit{OralCam} could give reasonable results for in-the-wild application, with an average ratings of 4.2 and 4.8, respectively.  
Also, they both mentioned that \textit{OralCam} performed accurately for photos taken with sufficient lighting and correct focus. 

Besides the lighting and focus issues as mentioned before, experts have provided more insights into other reasons for the model to make a mistake. 
For example, E1 mentioned that there was one photo focusing at the frontal part of cavity, and the side regions of cavity under shadow was wrongly flagged for soft deposit. 
To solve such problem, E1 suggested always taking multiple views of a cavity.
E2 mentioned that for three photos, the conditions were successfully detected, the flagged regions are questionable. 
In specific, for periodontal diseases, "\textit{... those at nipple are mostly correctly highlighted, but some at edge are not.}"

\section{8. Discussion}
In this section, we outline current limitations and discuss possible solutions for future work. 

\subsection{Smartphone Camera as Oral Sensor}
Smartphone cameras have enabled users to conveniently capture oral photos by themselves as input to \textit{OralCam}. 
However, we have noticed from in-the-wild application during the user study that two issues of photos can lead to detection errors:
{\em (i)}~ photos not focused correctly on oral cavity, and
{\em (ii)}~ noises added by improper lighting condition.
To be specific, the incorrect focus can result in very low resolution of photos, which has affected the detection of soft deposit in our study; 
and improper lighting can cause light spots on gingiva, which has confused the detection model as a sign of periodontal disease. 
We suggest several methods can be tried to solve in future work. 
For example, users can be clearly reminded to maintain the correct focus, and set thresholds of lighting for acceptable photos. 
Moreover, to further reduce false positive predictions, detecting and inpainting \cite{bertalmio2000image} of glare spots from photos can be applied.
However, the technique might not be sufficient for recovering the missing visual information of images.
In addition, real-time image quality feedback mechanisms as suggested by \cite{de2014bilicam} also seems promising for camera-based mHealth applications.

\subsection{Examination of Inner Cavity}
One of our current limitations is that the inner oral cavity (\textit{i.e.} occlusal surface of posterior teeth, lingual surface of teeth and gingiva) is not covered for examination due to a lack of proper lighting for capturing qualified photos.  
However, expert interviews have informed us the importance of information from that part of cavity for reaching comprehensive examination results. 
For example, caries most commonly exist at occlusal surface of posterior teeth for adults; 
and calculus are more obvious for accurate detection at lingual surface of teeth. 
Thus, we argue that capturing photos of inner cavity can possibly lead to more comprehensive and reliable exam results. 
Promising solutions may include utilizing emerging frontal flash of smartphone and designing accessories to provide sufficient lighting for photographing.  

\subsection{Gaining Trust from Users}
In \textit{OralCam}, detection results are presented with localization of the detected conditions. Visualizations, \textit{e.g.} the heatmap indicating where the model is `looking at', was specifically designed for gaining trust from layman users. 
However, our user study has shown that some participants still remained skeptical about \textit{OralCam}'s results. 
For example, one participant mentioned that, besides showing the location of conditions, it would help him to understand, and further trust the results, by giving reasons why the regions are flagged out. 
To be specific, he gave an example by saying: "\textit{it helps if it (\textit{OralCam}) can explain to me there are caries, because here is a black line-shape notch.}"
As such, recent progress of explainable DL \cite{zhang2018visual}, which aims to unbox DL model by interpreting the deep features learnt by the model, might become a promising solution.
Besides, external evidences, \textit{e.g.} FDA approvals and more comprehensive clinical trials, can be investigated for gaining user's trusts.

\section{9. Conclusion}
We have presented \textit{OralCam}, the first mobile app that enables self-examination of oral health for end users.  
\textit{OralCam} utilizes smartphone-captured oral photos and input from users' interaction for accurate detection of five common oral conditions.  
The detection results produced by deep learning are visualized hierarchically, probabilistically with localization to increase understandability for end users. 
\balance{}

\balance{}

\bibliographystyle{SIGCHI-Reference-Format}

\begin{thebibliography}{00}


\ifx \showCODEN    \undefined \def \showCODEN     #1{\unskip}     \fi
\ifx \showDOI      \undefined \def \showDOI       #1{{\tt DOI:}\penalty0{#1}\ }
  \fi
\ifx \showISBNx    \undefined \def \showISBNx     #1{\unskip}     \fi
\ifx \showISBNxiii \undefined \def \showISBNxiii  #1{\unskip}     \fi
\ifx \showISSN     \undefined \def \showISSN      #1{\unskip}     \fi
\ifx \showLCCN     \undefined \def \showLCCN      #1{\unskip}     \fi
\ifx \shownote     \undefined \def \shownote      #1{#1}          \fi
\ifx \showarticletitle \undefined \def \showarticletitle #1{#1}   \fi
\ifx \showURL      \undefined \def \showURL       #1{#1}          \fi

\bibitem{akther2019moral}
{Sayma Akther}, {Nazir Saleheen}, {Shahin~Alan Samiei}, {Vivek Shetty}, {Emre
  Ertin}, {and} {Santosh Kumar}. 2019.
\newblock \showarticletitle{mORAL: An mHealth Model for Inferring Oral Hygiene
  Behaviors in-the-wild Using Wrist-worn Inertial Sensors}.
\newblock {\em Proceedings of the ACM on Interactive, Mobile, Wearable and
  Ubiquitous Technologies\/} {3}, 1 (2019), 1.
\newblock


\bibitem{ash1964correlation}
{MM Ash}, {BN Gitlin}, {and} {WA Smith}. 1964.
\newblock \showarticletitle{Correlation between plaque and gingivitis}.
\newblock {\em The Journal of Periodontology\/} {35}, 5 (1964), 424--429.
\newblock


\bibitem{bertalmio2000image}
{Marcelo Bertalmio}, {Guillermo Sapiro}, {Vincent Caselles}, {and} {Coloma
  Ballester}. 2000.
\newblock \showarticletitle{Image inpainting}. In {\em Proceedings of the 27th
  annual conference on Computer graphics and interactive techniques}. ACM
  Press/Addison-Wesley Publishing Co., 417--424.
\newblock


\bibitem{broadbent2011dental}
{Jonathan~M Broadbent}, {W~Murray Thomson}, {John~V Boyens}, {and} {Richie
  Poulton}. 2011.
\newblock \showarticletitle{Dental plaque and oral health during the first 32
  years of life}.
\newblock {\em The Journal of the American Dental Association\/} {142}, 4
  (2011), 415--426.
\newblock


\bibitem{chang2008playful}
{Yu-Chen Chang}, {Jin-Ling Lo}, {Chao-Ju Huang}, {Nan-Yi Hsu}, {Hao-Hua Chu},
  {Hsin-Yen Wang}, {Pei-Yu Chi}, {and} {Ya-Lin Hsieh}. 2008.
\newblock \showarticletitle{Playful toothbrush: ubicomp technology for teaching
  tooth brushing to kindergarten children}. In {\em Proceedings of the SIGCHI
  conference on human factors in computing systems}. ACM, 363--372.
\newblock


\bibitem{cheng2016computer}
{Jie-Zhi Cheng}, {Dong Ni}, {Yi-Hong Chou}, {Jing Qin}, {Chui-Mei Tiu},
  {Yeun-Chung Chang}, {Chiun-Sheng Huang}, {Dinggang Shen}, {and} {Chung-Ming
  Chen}. 2016.
\newblock \showarticletitle{Computer-aided diagnosis with deep learning
  architecture: applications to breast lesions in US images and pulmonary
  nodules in CT scans}.
\newblock {\em Scientific reports\/}  {6} (2016), 24454.
\newblock


\bibitem{cicero2017training}
{Mark Cicero}, {Alexander Bilbily}, {Errol Colak}, {Tim Dowdell}, {Bruce Gray},
  {Kuhan Perampaladas}, {and} {Joseph Barfett}. 2017.
\newblock \showarticletitle{Training and validating a deep convolutional neural
  network for computer-aided detection and classification of abnormalities on
  frontal chest radiographs}.
\newblock {\em Investigative radiology\/} {52}, 5 (2017), 281--287.
\newblock


\bibitem{clawson2015no}
{James Clawson}, {Jessica~A Pater}, {Andrew~D Miller}, {Elizabeth~D Mynatt},
  {and} {Lena Mamykina}. 2015.
\newblock \showarticletitle{No longer wearing: investigating the abandonment of
  personal health-tracking technologies on craigslist}. In {\em Proceedings of
  the 2015 ACM International Joint Conference on Pervasive and Ubiquitous
  Computing}. ACM, 647--658.
\newblock


\bibitem{cohen2013expanding}
{Leonard~A Cohen}. 2013.
\newblock \showarticletitle{Expanding the physician's role in addressing the
  oral health of adults}.
\newblock {\em American journal of public health\/} {103}, 3 (2013), 408--412.
\newblock


\bibitem{de2018clinically}
{Jeffrey De~Fauw}, {Joseph~R Ledsam}, {Bernardino Romera-Paredes}, {Stanislav
  Nikolov}, {Nenad Tomasev}, {Sam Blackwell}, {Harry Askham}, {Xavier Glorot},
  {Brendan O'Donoghue}, {Daniel Visentin}, {and} {others}. 2018.
\newblock \showarticletitle{Clinically applicable deep learning for diagnosis
  and referral in retinal disease}.
\newblock {\em Nature medicine\/} {24}, 9 (2018), 1342.
\newblock


\bibitem{de2014bilicam}
{Lilian De~Greef}, {Mayank Goel}, {Min~Joon Seo}, {Eric~C Larson}, {James~W
  Stout}, {James~A Taylor}, {and} {Shwetak~N Patel}. 2014.
\newblock \showarticletitle{Bilicam: using mobile phones to monitor newborn
  jaundice}. In {\em Proceedings of the 2014 ACM International Joint Conference
  on Pervasive and Ubiquitous Computing}. ACM, 331--342.
\newblock


\bibitem{de2009periodontitis}
{Paola De~Pablo}, {Iain~LC Chapple}, {Christopher~D Buckley}, {and} {Thomas
  Dietrich}. 2009.
\newblock \showarticletitle{Periodontitis in systemic rheumatic diseases}.
\newblock {\em Nature Reviews Rheumatology\/} {5}, 4 (2009), 218.
\newblock


\bibitem{de2009accuracy}
{M De~Vrijer}, {WP Medendorp}, {and} {JAM Van~Gisbergen}. 2009.
\newblock \showarticletitle{Accuracy-precision trade-off in visual orientation
  constancy}.
\newblock {\em Journal of vision\/} {9}, 2 (2009), 9--9.
\newblock


\bibitem{deep2000screening}
{Paul Deep}. 2000.
\newblock \showarticletitle{Screening for common oral diseases}.
\newblock {\em Journal-Canadian Dental Association\/} {66}, 6 (2000), 298--299.
\newblock


\bibitem{ding2019reading}
{Xianghua Ding}, {Yanqi Jiang}, {Xiankang Qin}, {Yunan Chen}, {Wenqiang Zhang},
  {and} {Lizhe Qi}. 2019.
\newblock \showarticletitle{Reading Face, Reading Health: Exploring Face
  Reading Technologies for Everyday Health}. In {\em Proceedings of the 2019
  CHI Conference on Human Factors in Computing Systems}. ACM, 205.
\newblock


\bibitem{dou2016automatic}
{Qi Dou}, {Hao Chen}, {Lequan Yu}, {Lei Zhao}, {Jing Qin}, {Defeng Wang},
  {Vincent~CT Mok}, {Lin Shi}, {and} {Pheng-Ann Heng}. 2016.
\newblock \showarticletitle{Automatic detection of cerebral microbleeds from MR
  images via 3D convolutional neural networks}.
\newblock {\em IEEE transactions on medical imaging\/} {35}, 5 (2016),
  1182--1195.
\newblock


\bibitem{eke2012prevalence}
{Paul~I Eke}, {BA Dye}, {Li Wei}, {GO Thornton-Evans}, {and} {RJ Genco}. 2012.
\newblock \showarticletitle{Prevalence of periodontitis in adults in the United
  States: 2009 and 2010}.
\newblock {\em Journal of dental research\/} {91}, 10 (2012), 914--920.
\newblock


\bibitem{esteva2017dermatologist}
{Andre Esteva}, {Brett Kuprel}, {Roberto~A Novoa}, {Justin Ko}, {Susan~M
  Swetter}, {Helen~M Blau}, {and} {Sebastian Thrun}. 2017.
\newblock \showarticletitle{Dermatologist-level classification of skin cancer
  with deep neural networks}.
\newblock {\em Nature\/} {542}, 7639 (2017), 115.
\newblock


\bibitem{esteva2019guide}
{Andre Esteva}, {Alexandre Robicquet}, {Bharath Ramsundar}, {Volodymyr
  Kuleshov}, {Mark DePristo}, {Katherine Chou}, {Claire Cui}, {Greg Corrado},
  {Sebastian Thrun}, {and} {Jeff Dean}. 2019.
\newblock \showarticletitle{A guide to deep learning in healthcare}.
\newblock {\em Nature medicine\/} {25}, 1 (2019), 24.
\newblock


\bibitem{featherstone2008dental}
{JDB Featherstone}. 2008.
\newblock \showarticletitle{Dental caries: a dynamic disease process}.
\newblock {\em Australian dental journal\/} {53}, 3 (2008), 286--291.
\newblock


\bibitem{hashemian2015t}
{Tony~S Hashemian}, {Donna Kritz-Silverstein}, {and} {Ryan Baker}. 2015.
\newblock \showarticletitle{T ext2 F loss: the feasibility and acceptability of
  a text messaging intervention to improve oral health behavior and knowledge}.
\newblock {\em Journal of public health dentistry\/} {75}, 1 (2015), 34--41.
\newblock


\bibitem{hattab1999dental}
{Faiez~N Hattab}, {Muawia~A Qudeimat}, {and} {Hala~S Al-Rimawi}. 1999.
\newblock \showarticletitle{Dental discoloration: an overview}.
\newblock {\em Journal of Esthetic and Restorative Dentistry\/} {11}, 6 (1999),
  291--310.
\newblock


\bibitem{heymann2005tooth}
{HO Heymann}. 2005.
\newblock \showarticletitle{Tooth whitening: facts and fallacies}.
\newblock {\em British Dental Journal\/} {198}, 8 (2005), 514.
\newblock


\bibitem{huang2016toothbrushing}
{Hua Huang} {and} {Shan Lin}. 2016.
\newblock \showarticletitle{Toothbrushing monitoring using wrist watch}. In
  {\em Proceedings of the 14th ACM Conference on Embedded Network Sensor
  Systems CD-ROM}. ACM, 202--215.
\newblock


\bibitem{jepsen2011calculus}
{S{\o}ren Jepsen}, {James Deschner}, {Andreas Braun}, {Frank Schwarz}, {and}
  {J{\"o}rg Eberhard}. 2011.
\newblock \showarticletitle{Calculus removal and the prevention of its
  formation}.
\newblock {\em Periodontology 2000\/} {55}, 1 (2011), 167--188.
\newblock


\bibitem{jiang2017artificial}
{Fei Jiang}, {Yong Jiang}, {Hui Zhi}, {Yi Dong}, {Hao Li}, {Sufeng Ma}, {Yilong
  Wang}, {Qiang Dong}, {Haipeng Shen}, {and} {Yongjun Wang}. 2017.
\newblock \showarticletitle{Artificial intelligence in healthcare: past,
  present and future}.
\newblock {\em Stroke and vascular neurology\/} {2}, 4 (2017), 230--243.
\newblock


\bibitem{joybell2015comparison}
{Chrishantha Joybell}, {Ramesh Krishnan}, {and} {Suresh Kumar}. 2015.
\newblock \showarticletitle{Comparison of Two Brushing Methods-Fone’s vs
  Modified Bass Method in Visually Impaired Children Using the Audio Tactile
  Performance (ATP) Technique}.
\newblock {\em Journal of clinical and diagnostic research: JCDR\/} {9}, 3
  (2015), ZC19.
\newblock


\bibitem{kassebaum2017global}
{NJ Kassebaum}, {AGC Smith}, {E Bernab{\'e}}, {TD Fleming}, {AE Reynolds}, {T
  Vos}, {CJL Murray}, {W Marcenes}, {and} {GBD 2015 Oral~Health Collaborators}.
  2017.
\newblock \showarticletitle{Global, regional, and national prevalence,
  incidence, and disability-adjusted life years for oral conditions for 195
  countries, 1990--2015: a systematic analysis for the global burden of
  diseases, injuries, and risk factors}.
\newblock {\em Journal of dental research\/} {96}, 4 (2017), 380--387.
\newblock


\bibitem{korpela2015evaluating}
{Joseph Korpela}, {Ryosuke Miyaji}, {Takuya Maekawa}, {Kazunori Nozaki}, {and}
  {Hiroo Tamagawa}. 2015.
\newblock \showarticletitle{Evaluating tooth brushing performance with
  smartphone sound data}. In {\em Proceedings of the 2015 ACM International
  Joint Conference on Pervasive and Ubiquitous Computing}. ACM, 109--120.
\newblock


\bibitem{litjens2017survey}
{Geert Litjens}, {Thijs Kooi}, {Babak~Ehteshami Bejnordi}, {Arnaud
  Arindra~Adiyoso Setio}, {Francesco Ciompi}, {Mohsen Ghafoorian}, {Jeroen~Awm
  Van Der~Laak}, {Bram Van~Ginneken}, {and} {Clara~I S{\'a}nchez}. 2017.
\newblock \showarticletitle{A survey on deep learning in medical image
  analysis}.
\newblock {\em Medical image analysis\/}  {42} (2017), 60--88.
\newblock


\bibitem{liu2016ssd}
{Wei Liu}, {Dragomir Anguelov}, {Dumitru Erhan}, {Christian Szegedy}, {Scott
  Reed}, {Cheng-Yang Fu}, {and} {Alexander~C Berg}. 2016.
\newblock \showarticletitle{Ssd: Single shot multibox detector}. In {\em
  European conference on computer vision}. Springer, 21--37.
\newblock


\bibitem{loesche2001periodontal}
{Walter~J Loesche} {and} {Natalie~S Grossman}. 2001.
\newblock \showarticletitle{Periodontal disease as a specific, albeit chronic,
  infection: diagnosis and treatment}.
\newblock {\em Clinical microbiology reviews\/} {14}, 4 (2001), 727--752.
\newblock


\bibitem{mariakakis2017biliscreen}
{Alex Mariakakis}, {Megan~A Banks}, {Lauren Phillipi}, {Lei Yu}, {James
  Taylor}, {and} {Shwetak~N Patel}. 2017a.
\newblock \showarticletitle{Biliscreen: smartphone-based scleral jaundice
  monitoring for liver and pancreatic disorders}.
\newblock {\em Proceedings of the ACM on Interactive, Mobile, Wearable and
  Ubiquitous Technologies\/} {1}, 2 (2017), 20.
\newblock


\bibitem{mariakakis2017pupilscreen}
{Alex Mariakakis}, {Jacob Baudin}, {Eric Whitmire}, {Vardhman Mehta}, {Megan~A
  Banks}, {Anthony Law}, {Lynn Mcgrath}, {and} {Shwetak~N Patel}. 2017b.
\newblock \showarticletitle{PupilScreen: Using Smartphones to Assess Traumatic
  Brain Injury}.
\newblock {\em Proceedings of the ACM on Interactive, Mobile, Wearable and
  Ubiquitous Technologies\/} {1}, 3 (2017), 81.
\newblock


\bibitem{mariotti1999dental}
{Angelo Mariotti}. 1999.
\newblock \showarticletitle{Dental plaque-induced gingival diseases}.
\newblock {\em Annals of periodontology\/} {4}, 1 (1999), 7--17.
\newblock


\bibitem{nasseh2014effect}
{Kamyar Nasseh}, {Barbara Greenberg}, {Marko Vujicic}, {and} {Michael Glick}.
  2014.
\newblock \showarticletitle{The effect of chairside chronic disease screenings
  by oral health professionals on health care costs}.
\newblock {\em American journal of public health\/} {104}, 4 (2014), 744--750.
\newblock


\bibitem{nazir2017prevalence}
{Muhammad~Ashraf Nazir}. 2017.
\newblock \showarticletitle{Prevalence of periodontal disease, its association
  with systemic diseases and prevention}.
\newblock {\em International journal of health sciences\/} {11}, 2 (2017), 72.
\newblock


\bibitem{ouyang2017asymmetrical}
{Zhenchao Ouyang}, {Jingfeng Hu}, {Jianwei Niu}, {and} {Zhiping Qi}. 2017.
\newblock \showarticletitle{An asymmetrical acoustic field detection system for
  daily tooth brushing monitoring}. In {\em GLOBECOM 2017-2017 IEEE Global
  Communications Conference}. IEEE, 1--6.
\newblock


\bibitem{parker2019availability}
{Kate Parker}, {Rozana~Valiji Bharmal}, {and} {Mohammad~Owaise Sharif}. 2019.
\newblock \showarticletitle{The availability and characteristics of
  patient-focused oral hygiene apps}.
\newblock {\em British dental journal\/} {226}, 8 (2019), 600.
\newblock


\bibitem{petersen2005global}
{Poul~Erik Petersen}, {Denis Bourgeois}, {Hiroshi Ogawa}, {Saskia
  Estupinan-Day}, {and} {Charlotte Ndiaye}. 2005.
\newblock \showarticletitle{The global burden of oral diseases and risks to
  oral health}.
\newblock {\em Bulletin of the World Health Organization\/}  {83} (2005),
  661--669.
\newblock


\bibitem{ranjan2017all}
{Rajeev Ranjan}, {Swami Sankaranarayanan}, {Carlos~D Castillo}, {and} {Rama
  Chellappa}. 2017.
\newblock \showarticletitle{An all-in-one convolutional neural network for face
  analysis}. In {\em 2017 12th IEEE International Conference on Automatic Face
  \& Gesture Recognition (FG 2017)}. IEEE, 17--24.
\newblock


\bibitem{redmon2017yolo9000}
{Joseph Redmon} {and} {Ali Farhadi}. 2017.
\newblock \showarticletitle{YOLO9000: better, faster, stronger}. In {\em
  Proceedings of the IEEE conference on computer vision and pattern
  recognition}. 7263--7271.
\newblock


\bibitem{reisman2017ehrs}
{Miriam Reisman}. 2017.
\newblock \showarticletitle{EHRs: the challenge of making electronic data
  usable and interoperable}.
\newblock {\em Pharmacy and Therapeutics\/} {42}, 9 (2017), 572.
\newblock


\bibitem{selvaraju2017grad}
{Ramprasaath~R Selvaraju}, {Michael Cogswell}, {Abhishek Das}, {Ramakrishna
  Vedantam}, {Devi Parikh}, {and} {Dhruv Batra}. 2017.
\newblock \showarticletitle{Grad-cam: Visual explanations from deep networks
  via gradient-based localization}. In {\em Proceedings of the IEEE
  International Conference on Computer Vision}. 618--626.
\newblock


\bibitem{tiffany2018mobile}
{Brooks Tiffany}, {Paula Blasi}, {Sheryl~L Catz}, {and} {Jennifer~B McClure}.
  2018.
\newblock \showarticletitle{Mobile apps for oral health promotion: content
  review and heuristic usability analysis}.
\newblock {\em JMIR mHealth and uHealth\/} {6}, 9 (2018), e11432.
\newblock


\bibitem{underwood2015use}
{Ben Underwood}, {J Birdsall}, {and} {E Kay}. 2015.
\newblock \showarticletitle{The use of a mobile app to motivate evidence-based
  oral hygiene behaviour}.
\newblock {\em British dental journal\/} {219}, 4 (2015), E2.
\newblock


\bibitem{vardell2012visualdx}
{Emily Vardell} {and} {Carmen Bou-Crick}. 2012.
\newblock \showarticletitle{VisualDx: a visual diagnostic decision support
  tool}.
\newblock {\em Medical reference services quarterly\/} {31}, 4 (2012),
  414--424.
\newblock


\bibitem{wang2018seismo}
{Edward~Jay Wang}, {Junyi Zhu}, {Mohit Jain}, {Tien-Jui Lee}, {Elliot Saba},
  {Lama Nachman}, {and} {Shwetak~N Patel}. 2018b.
\newblock \showarticletitle{Seismo: Blood pressure monitoring using built-in
  smartphone accelerometer and camera}. In {\em Proceedings of the 2018 CHI
  Conference on Human Factors in Computing Systems}. ACM, 425.
\newblock


\bibitem{wang2018tienet}
{Xiaosong Wang}, {Yifan Peng}, {Le Lu}, {Zhiyong Lu}, {and} {Ronald~M Summers}.
  2018a.
\newblock \showarticletitle{Tienet: Text-image embedding network for common
  thorax disease classification and reporting in chest x-rays}. In {\em
  Proceedings of the IEEE conference on computer vision and pattern
  recognition}. 9049--9058.
\newblock


\bibitem{xu2018ecglens}
{Ke Xu}, {Shunan Guo}, {Nan Cao}, {David Gotz}, {Aiwen Xu}, {Huamin Qu},
  {Zhenjie Yao}, {and} {Yixin Chen}. 2018a.
\newblock \showarticletitle{Ecglens: Interactive visual exploration of large
  scale ecg data for arrhythmia detection}. In {\em Proceedings of the 2018 CHI
  Conference on Human Factors in Computing Systems}. ACM, 663.
\newblock


\bibitem{xu2018less}
{Zhoubing Xu}, {Yuankai Huo}, {JinHyeong Park}, {Bennett Landman}, {Andy
  Milkowski}, {Sasa Grbic}, {and} {Shaohua Zhou}. 2018b.
\newblock \showarticletitle{Less is more: Simultaneous view classification and
  landmark detection for abdominal ultrasound images}. In {\em International
  Conference on Medical Image Computing and Computer-Assisted Intervention}.
  Springer, 711--719.
\newblock


\bibitem{yoshitani2016lumio}
{Takuma Yoshitani}, {Masa Ogata}, {and} {Koji Yatani}. 2016.
\newblock \showarticletitle{LumiO: a plaque-aware toothbrush}. In {\em
  Proceedings of the 2016 ACM International Joint Conference on Pervasive and
  Ubiquitous Computing}. ACM, 605--615.
\newblock


\bibitem{zhang2018visual}
{Quan-shi Zhang} {and} {Song-Chun Zhu}. 2018.
\newblock \showarticletitle{Visual interpretability for deep learning: a
  survey}.
\newblock {\em Frontiers of Information Technology \& Electronic Engineering\/}
  {19}, 1 (2018), 27--39.
\newblock


\bibitem{zhang2017mdnet}
{Zizhao Zhang}, {Yuanpu Xie}, {Fuyong Xing}, {Mason McGough}, {and} {Lin Yang}.
  2017.
\newblock \showarticletitle{Mdnet: A semantically and visually interpretable
  medical image diagnosis network}. In {\em Proceedings of the IEEE conference
  on computer vision and pattern recognition}. 6428--6436.
\newblock


\bibitem{zhou2016learning}
{Bolei Zhou}, {Aditya Khosla}, {Agata Lapedriza}, {Aude Oliva}, {and} {Antonio
  Torralba}. 2016.
\newblock \showarticletitle{Learning deep features for discriminative
  localization}. In {\em Proceedings of the IEEE conference on computer vision
  and pattern recognition}. 2921--2929.
\newblock


\end{thebibliography}

\end{document}